%% file: paper.tex
\documentclass[sigconf, review]{acmart}

\renewcommand\footnotetextcopyrightpermission[1]{} 

\usepackage{hyperref}

\usepackage{tabularx,adjustbox}
\usepackage{booktabs}
\usepackage{graphicx}  
\usepackage{algorithm,algpseudocode} 
\usepackage{longfbox}
\usepackage{subcaption}
\usepackage{tcolorbox}

\usepackage{amsmath, amscd, mathrsfs}

\usepackage{multirow}

\usepackage{array}
\usepackage{url}

\usepackage{pifont}

\makeatletter
\newcommand{\vast}{\bBigg@{4}}
\newcommand{\Vast}{\bBigg@{5}}
\newcommand{\immense}{\bBigg@{6}}
\makeatother

\usepackage{ragged2e}

\begin{document}

\title{EPhishCADE: A Privacy-Aware Multi-Dimensional Framework \\ for Email Phishing Campaign Detection}

\author{Wei Kang}
\email{kangism@outlook.com}
\affiliation{
\institution{Gloglo Technology Ltd.}
\city{}
\state{}
\country{Canada}
}

\author{Nan Wang}
\email{nan.wang@data61.csiro.au}
\affiliation{
\institution{CSIRO's Data61}
\city{}
\state{}
\country{Australia}
}

\author{Jang Seung}
\email{seung.jang@data61.csiro.au}
\affiliation{
\institution{CSIRO's Data61}
\city{}
\state{}
\country{Australia}
}

\author{Shuo Wang}
\email{wangshuosj@sjtu.edu.cn}
\affiliation{
\institution{Shanghai Jiao Tong University}
\city{}
\state{}
\country{China}
}

\author{Alsharif Abuadbba}
\email{sharif.abuadbba@data61.csiro.au}
\affiliation{
\institution{CSIRO's Data61}
\city{}
\state{}
\country{Australia}
}

\begin{abstract}
\input{abstract}
\end{abstract}

\keywords{Email Phishing Campaigns, Clustering, Graph-based Detection, Multi-Dimension, Privacy-Awareness, Cyber Intelligence Analytics}

\maketitle

\pagestyle{plain} 

\input{intro}
\input{problem}

\input{intuition}
\input{algo}

\input{experiment1}

\input{experiment}

\input{conclusion}

\bibliographystyle{ACM-Reference-Format}
\bibliography{reference}

\end{document}

%% file: abstract.tex
Phishing attacks, typically carried out by email, remain a significant cybersecurity threat with attackers creating legitimate-looking websites to deceive recipients into revealing sensitive information or executing harmful actions. Beyond traditional phishing attack detection, organizations face new challenges that pose significant economic and security risks: 1) a rise in coordinated phishing attacks that are more likely to succeed; 2) limited resources to manage a flood of phishing alerts and reports, leading to delays or oversights in addressing critical threats; and 3) increased privacy and security concerns regarding sensitive personal or national information, especially within government agencies.

To address these emergent challenges, we propose {\bf EPhishCADE}, the first {\em privacy-aware}, {\em multi-dimensional} framework for {\bf E}mail {\bf Phish}ing {\bf CA}mpaign {\bf DE}tection to automatically identify email phishing campaigns by clustering seemingly unrelated attacks. Our framework employs a hierarchical architecture combining a structural layer and a contextual layer, offering a comprehensive analysis of phishing attacks by thoroughly examining both structural and contextual elements. Specifically, we implement a graph-based contextual layer to reveal hidden similarities across multiple dimensions, including textual, numeric, temporal, and spatial features, among attacks that may initially appear unrelated. Our framework streamlines the handling of security threat reports, reducing analysts' fatigue and workload while enhancing protection against these threats. Another key feature of our framework lies in its sole reliance on phishing URLs in emails without the need for private information, including senders, recipients, content, etc. This feature enables a collaborative identification of phishing campaigns and attacks among multiple organizations without compromising privacy. Finally, we benchmark our framework against an established structure-based study (WWW \textquotesingle 17) to demonstrate its effectiveness.

%% file: intro.tex
\section{Introduction} \label{sec:intro}

Phishing attacks remain one of the most prevalent and insidious forms of cybersecurity threats faced by individuals and organizations worldwide. These attacks, typically carried out via emails, aim to deceive recipients into divulging sensitive information, such as login credentials, financial information, or other personal data, by masquerading as trustworthy entities. Phishing emails generally contain malicious links leading to fake websites that closely resemble legitimate ones, designed to capture sensitive information of victims. Organizations generally rely on two main ways for collecting phishing emails. One major approach involves using various security tools and systems to automatically scan incoming emails and generate alerts when potential phishing attacks are detected. However, due to the evolving nature of phishing tactics, where attackers continually refine their methods to evade detection, manual reporting remains an essential complement. Employees tend to be encouraged to report suspected phishing emails that might bypass security tools. 

\subsection{Challenges}

Currently, most existing studies have concentrated on detecting email phishing attacks by employing a variety of techniques and methodologies, such as machine learning-based approaches \cite{GrantHo2019Detecting,li2020lstm,alam2020phishing,smadi2018detection,form2015phishing, clusterblacklist, sabir2022reliability,evans2022raider}, behavioral analysis \cite{hakim2021phishing,abroshan2021phishing,kashapov2022email,shmalko2022profiler}, natural language processing (NLP) techniques \cite{salloum2022systematic,alhogail2021applying,verma2020email,haynes2021lightweight}, ensemble methods \cite{basit2020novel} and graph-based approach \cite{graphdetection}. These efforts primarily focus on classifying emails as legitimate or phishing, making it increasingly difficult for attackers to succeed with isolated attacks. However, these approaches often overlook the latent connections between different phishing attacks and the broader phishing campaigns involved, leaving organizations exposed to emerging challenges:
\begin{itemize}
    \item{\bf Coordinated Attacks.} Phishing attacks tend to be performed in a coordinated way to maximize success rates. Attackers carefully design and carry out coordinated sets of attacks, often referred to as campaigns, leveraging in-depth knowledge of their targets' behaviors and systems. For instance, they may send out multiple waves of seemingly legitimate emails in different templates to mimic various targets. By employing sophisticated social engineering tactics, attackers can manipulate recipients into taking harmful actions, all while avoiding detection by security tools designed to block such threats. These well-orchestrated campaigns blend technical expertise with psychological manipulation, making them particularly challenging to identify and defend against. 

    \item{\bf Limited Resources.} Security analysts face a daily influx of overwhelming alerts and reports from cybersecurity monitoring systems, including a large number of emails alerts, requiring them to carefully process them through a series of key steps, including identifying threats and investigating attack vectors to determine the extent of the breach, and implementing mitigation measures. This heavy workload not only diverts their attention from crucial security issues but also leads to fatigue, causing critical security threats to be overlooked or not addressed promptly, exposing organizations to significant economic loss ans security risks.

    \item{\bf Privacy Dilemma.} Organizations, especially government agencies, also face a significant privacy dilemma. On the one hand, they must protect sensitive personal or national information such as email contents, sender and recipient details, and other potentially identifiable data. While sharing alerts and reports with untrusted third-party services, like cloud providers, could help detect malicious content, on the contrary it introduces risks of privacy breaches and potential attack vectors. On the other hand, operating in isolation can effectively prevent privacy leakage, but it limits the ability to harness collective intelligence from collaborative efforts. This isolation can lead to missed opportunities in recognizing broader attack patterns or evolving tactics that could be identified through shared data. As a result, organizations become more vulnerable and less capable of responding effectively to sophisticated and widespread phishing campaigns.
\end{itemize}
This leaves us an intriguing and challenging question:
\begin{tcolorbox}[colback=red!5!white,colframe=red!75!black,title=Question]
Is there a comprehensive solution to effectively address the multifaceted challenges posed by coordinated attacks, limited resources and the privacy dilemma?
\end{tcolorbox}

\begin{tcolorbox}[colback=blue!5!white,colframe=blue!75!black,title=Contribution]
We propose the first {\em privacy-aware}, {\em multi-dimensional} framework, {\bf EPhishCADE}, ({\bf E}mail {\bf Phish}ing {\bf CA}mpaign {\bf DE}tection) to automatically identify email phishing campaigns. It clusters seemingly unrelated phishing attacks by uncovering latent multi-dimensional connections among them, including textual, numeric, temporal, and spatial features. By combining structural features with contextual information extracted from URLs in emails, this dual-layered approach improves the detection process, offering a deeper understanding of phishing tactics and reducing the number of campaigns, thereby alleviating workloads.
\end{tcolorbox}

\subsection{Motivation}

Recall that in phishing emails, attackers entice users to click on a URL link which directs them to a legitimate-looking website where sensitive information is requested. Thus, it becomes vital to analyze these URLs and their underlying relations to identify campaigns of phishing attacks~\cite{abuadbba2022towards,almashor2023unraveling,wang2023doitrust}. However, URLs themselves are often less informative and cannot be used solely for this purpose. For instance, attackers may use URL shorteners to generate random URLs to disguise the malicious ones so that the phishing URL looks benign. Moreover, existing methods predominantly rely on the structural information of webpages, such as layout patterns, to perform clustering for detecting phishing campaigns. While these approaches can achieve a certain level of effectiveness, they often fall short in fully capturing the nuanced strategies employed by attackers. By focusing solely on structural attributes, they miss the deeper semantic context, such as the language used in text, the intent behind visual elements, and the specific wording that lures victims, which are crucial for distinguishing sophisticated phishing attempts. This limitation reduces their ability to detect more complex or adaptive phishing tactics that evolve beyond simple structural patterns, highlighting the need for a more holistic approach that integrates both structural and contextual analysis. 

Thus, we come up with the idea of collecting and exploiting additional multi-dimensional contextual information, such as the associated IP address, the HTML content, and the screenshots of webpages, to enrich the contextual understanding of URLs. By incorporating these diverse data sources, our approach aims to enhance the detection of sophisticated phishing campaigns, capturing subtle cues that traditional methods often overlook. In our framework, we construct graphs to reveal latent and complex connections among multiple phishing URLs, as these URLs often exhibit shared patterns and features. This graph-based method has shown robustness in detecting phishing attacks \cite{graphdetection}, and we further investigate its potential to identify broader phishing campaigns rather than binary classification of individual phishing websites.

\subsection{Advancements}

Detection of phishing campaigns is a subclass of a clustering problem. It remains an open challenge with limited studies attempting to cluster phishing emails to find connections among phishing attacks. Most relevant studies focus on using sensitive information about emails without taking privacy into consideration. One line of work \cite{hamid2013profiling, hamid2014approach, DINH2015S12, Saka2022, althobaiti2023using} focuses on applying various clustering algorithms, e.g., K-means, DBSCAN, Meanshift, etc., to group phishing emails into campaigns based on multiple features associated with emails, including the details of time, subject, body, attachment, sender, recipient and url. 

Another popular line of research, exemplified by two notable studies, focuses on clustering using only URL-based details, which are relevant to our work. They proposed two distinct approaches for clustering phishing websites, both based on the same intuition that attackers often reuse templates when constructing phishing sites. The study \cite{md5cluster} focuses on clustering phishing websites that share a similar ratio of files with identical {\em MD5 hash values}. Specifically, their method calculates the MD5 hash values of all the files related to a phishing website, including HTML files, image files, css files, etc. Any two phishing websites are considered similar if the ratio is above a certain threshold. However, this approach is vulnerable to evasion, as attackers can avoid detection by making even a minor alteration, such as changing a single character in relevant files, which results in different MD5 hashes. The other study \cite{DOMCluster} groups phishing websites by similar DOM structures. The approach generates ``tag vectors'' that represent the frequency of HTML tags used in the websites. This method defines a {\em proportional distance} between the tag vectors of two phishing websites by dividing the number of differences by the total number of tags present in the websites. Two phishing websites are considered similar if their proportional distance falls below a certain threshold. These two solutions rely solely on a single structural dimension, making them unable to reveal deeper contextual connections among different phishing attacks.

\begin{table*} [t]
\caption{Example Phishing URLs and Signals}
\label{tab:example_malicious_urls}
\centering
\resizebox{0.76\textwidth}{!}{%
\begin{tabular}{c|c|c|c|c} 
   \hline
   \textbf{URLs} & \textbf{URL tokens}  & \textbf{Associated IPs} & \textbf{Target} & \textbf{HTML tokens} \\
   \hline
   $u_1$:s286.paypal-login.net & s286, paypal, login, net & \{168.10.10.2\} & Paypal & paypal, login, password \\
   \hline
   $u_2$:s8790.paypal-login.net & s8790, paypal, login, net & \{168.10.10.2\} & Paypal & paypal,log-in, passwd \\
   \hline
   $u_3$:aws-amazon.net.au & aws, amazon, net, au  & \{198.111.0.59\} & Other & amazon, aws\\
   \hline
\end{tabular}}
\end{table*}

\subsection{Benefits}

Generally, our framework allows to gain deep insights into detected campaigns enabling analysts to assess the scale, size, duration, urgency and complexity of these campaigns. With this comprehensive understanding, analysts can prioritize responses based on the assessment outcomes and proactively establish strategies to defend against future attacks with similar characteristics to known campaigns. It streamlines the processing and prioritization of security threat reports, significantly reducing analysts' fatigue and workload, and thereby maximizing protection against these threats. Compared to existing studies, our framework has two distinguishing features:  
\begin{itemize}    
    \item{\bf Privacy-Awareness:} Unlike privacy-unaware methods, our approach only relies on the URL-related details without the need for sensitive data about emails. This facilitates the collaborative identification of phishing campaigns across multiple organizations while preserving privacy.

    \item{\bf Multi-Dimension:} Traditional studies typically rely on structure-based, single-dimension approaches. In contrast, our method adopts a multi-dimensional strategy by combining both structural and contextual information. This enables more fine-grained identification of phishing campaigns from multiple perspectives, empowering analysts to develop targeted defense strategies against future threats. Additionally, the multi-dimensional approach significantly increases the difficulty for attackers to evade detection, enhancing overall phishing defense effectiveness.
\end{itemize}

\subsection{Outline of Our Paper}

Our paper is organized as follows. First, we formalize the problem in Section \ref{sec:problem}. We elaborate on the specifics of our framework in Section \ref{sec:stage1} and \ref{sec:stage2}, respectively. Finally, we benchmark our framework against an established structure-based study in Section \ref{sec:experiment}. 

%% file: problem.tex
\section{Problem Definition} \label{sec:problem}

\subsection{Preliminaries} \label{sec:preliminary}

In this section, we provide the crucial definitions:

\subsubsection{URL Instance} 

\begin{definition}[Phishing URL Instance]
A phishing URL instance, denoted by $u$, represents a specific instantiation of cyber threat where the URL is employed as a vector to launch attacks by malicious actors.  
\end{definition}

It is important to understand that the maliciousness of a URL is not inherent in its characters or structure but is derived from the intent behind its creation and use. This intent is manifested in several ways, such as directing users to harmful content, deceiving users into divulging sensitive information, or silently triggering unauthorized actions on users' devices. 

\subsubsection{Signal} 

\begin{definition}[Signal]
A signal $s_i$ ($0<i\leq|S|$) is a feature or attribute which provides additional information of a URL, where $S$ is a signal set and $|\cdot|$ indicates the cardinality of a set. Given a URL $u$, the value of signal $s_i$ is denoted by $s_i(u)$.
\end{definition}
Signals provide additional ``intelligence'' for a URL, such as domain names, lexical features, geolocations or temporal patterns. There are two main categories of signals -- texts and comparable values. For instance, the HTML content or OCR text extracted from a screenshot of a webpage falls into the first category. Text information is not directly comparable, but a number of techniques, such as TF-IDF \cite{qaiser2018text} and embeddings, \cite{wang2020comparative,selva2021review}
can be applied to calculate the similarity of textual signals if we want to compare two URLs. Other signals, such as IP address and DNS, are in the second category, where the values are directly comparable. To collect the values of selected signals for phishing URLs, we need a process called \textit{signal enrichment}, where the information of these signals is collected from one or more threat intelligence platforms, then integrated and stored in the database to facilitate further analysis. A URL is called an \textit{enriched URL} after additional signal information is collected. 

\subsubsection{Campaign} After enrichment information is collected, phishing URLs will be analyzed and grouped into campaigns. URLs in the same campaign are highly similar or related across multiple signals. This similarity or relevance manifests in multiple dimensions, with each signal representing a dimension. Although it is possible to observe discrepancies between the values of certain signals for two URLs in a campaign, the values of most other signals are expected to be highly similar.

\begin{definition}[Phishing Campaign]
A phishing campaign, denoted by $c$, constitutes a coordinated set of activities disseminating phishing URLs. This campaign is characterized by a collection of phishing URL instances that exhibit significant similarities or relevance w.r.t. a certain set $S$ of signals. 
\end{definition}

\begin{figure*}[!h]
\centering
\includegraphics[width=0.7\textwidth]{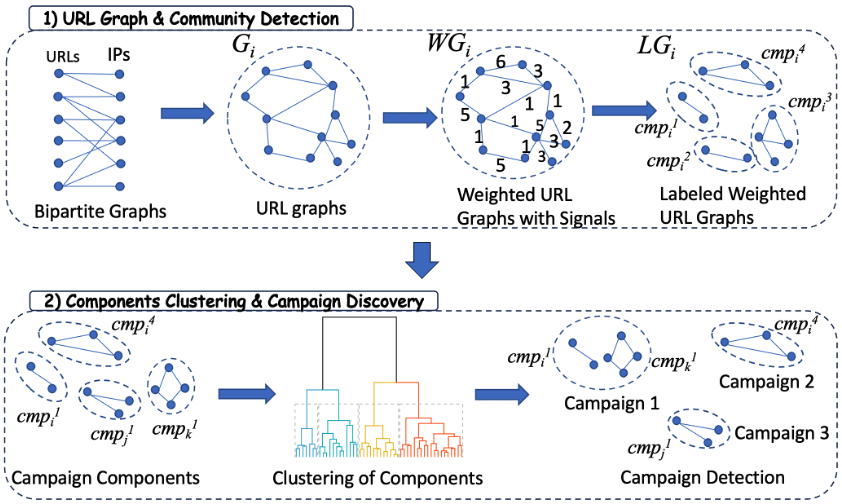}
\caption{The workflow of the contextual layer}
\label{fig:framework}
\end{figure*}

\begin{figure}[!h]
\centering
\includegraphics[width=0.35\textwidth]{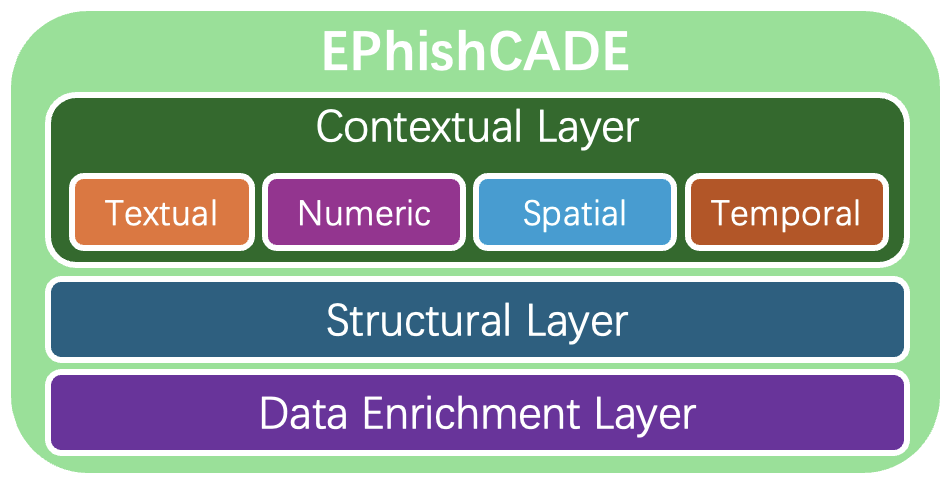}
\caption{The architecture of our framework}
\label{fig:architecture}
\end{figure}

The concept of a phishing campaign suggests a coordinated effort with a deliberate strategy behind the clustering of malicious URL instances. In our framework, we define a typical campaign as originating from a common source or targeting a common victim. Such a campaign may involve one attacker launching multiple phishing attacks against a single target brand, often leveraging the same underlying infrastructure, such as a shared physical server or a network of compromised systems, to execute the attacks.

The objectives of such campaigns can vary widely, ranging from financial gain, data theft, espionage, and disruption of services to broader strategic goals aligned with state-sponsored activities or large-scale fraud operations. For instance, there are three phishing URLs $u_1$, $u_2$, $u_3$ and four signals \texttt{URL tokens}, \texttt{Associated IPs}, \texttt{Target} and \texttt{HTML tokens} in Table \ref{tab:example_malicious_urls}. $u_1$ and $u_2$ have similar patterns in the URL strings, the same set of IP addresses associated with the URLs, the same target and very similar HTML tokens. Therefore, it is highly likely that $u_1$ and $u_2$ are two relevant phishing attack instances launched in the same organized phishing campaign attacking PayPal. On the contrary, $u_3$ is quite different from $u_1$ and $u_2$ across all the signals, and may belong to another campaign with a different attacking purpose. In this paper, we assume that all the URLs we deal with are phishing URLs and that each campaign consists of a number of these URLs. It is worth noting that, although we only consider phishing attacks, our proposed approach can be used to discover campaigns of any URL-based malicious attacks.

\subsection{Problem Definition}



\textbf{Detection of Phishing Campaigns}. 
Consider a set of phishing URLs denoted by $\mathcal{U}=\{u_1, u_2, ..., u_n\}$, where each URL $u_i$ represents a specific phishing URL instance. Accompanying these URLs is a set of signals $S=\{s_1, s_2, ..., s_{|S|}\}$, with each signal $s_j$ offering a specific attribute associated with the URLs.
The goal of this paper is to discover distinct phishing campaigns, denoted by $C=\{c_1, c_2, ..., c_m\}$, from the URL set $\mathcal{U}$ w.r.t. the signals in $S$. Each campaign $c_i$ ($1\leq i \leq m$) is an aggregation of URLs that share commonalities in their signal profiles, suggesting a coordinated effort in a series of relevant attacks. 



%% file: intuition.tex
\section{Intuition}

\subsection{Architecture} 

EPhishCADE employs an hierarchical architecture consisting of a data enrichment layer, a structural layer, and a contextual layer, as illustrated in Figure \ref{fig:architecture}. Phishing emails are processed through multiple layers in sequence. First, the data enrichment layer extracts URLs from the emails and enhances them with additional information. Then, the structural layer applies coarse clustering, drawing on the established insight \cite{DOMCluster} that attackers frequently reuse website layout templates. Lastly, by analyzing multi-dimensional contextual features, our framework refines the clustering, uncovering hidden links between phishing attacks with similar web structures, allowing for more precise detection and analysis.

\subsection{Data Enrichment} 

Phishing URLs are extracted from incoming alerts and reports without exposing any sensitive email information. These URLs are then enriched by collecting various attributes from third-party intelligence platforms, such as PhishTank, which provide additional context and insights about the URLs, enhancing the overall detection process.

\subsection{Structural Layer} The structural layer examines the core features that shape the layout of phishing websites, such as HTML tags, URL structures, and DNS information. By analyzing these elements, the structural layer identifies patterns that signify malicious intent. Understanding this layer is essential for phishing detection, as it captures the foundational blueprint that attackers manipulate to deceive users. Previous efforts, such as those in \cite{DOMCluster, md5cluster}, have demonstrated promising results in this area.

\subsection{Contextual Layer} We primarily focus on the multi-dimensional contextual layer. This layer enriches phishing URLs with multi-faceted contextual features gathered from third-party threat intelligence platforms. By analyzing various types of information embedded in these features, such as textual, numeric, spatial, temporal data, this layer builds weighted URL graphs to identify phishing campaigns. Phishing URLs with similar or relevant features can be clustered together to indicate some certain malicious attack. Fig. \ref{fig:framework} depicts the workflow of the contextual layer:
\begin{enumerate}
    \item At stage 1, bipartite graphs are constructed between URLs and their corresponding IP addresses. From these bipartite graphs, URL graphs are generated by adding edges between pairs of URLs that share a common IP address. These URL graphs are then transformed into weighted URL graphs, where the edge weights are assigned based on specific signals. Following this, weighted community detection is performed on the graphs to identify communities, with each community representing a subgraph of closely related URLs that are likely part of the same phishing campaign.

    \item At stage 2, each community is treated as an atomic unit, referred as component. These components are grouped into clusters based on their semantic similarity. A campaign can consist of one or more components, identified through the combination of structural and semantic information embedded in the signals.
\end{enumerate}

%% file: algo.tex
\section{Weighted URL Graph Detection}\label{sec:stage1}

In this section, we elaborate on the specifics of our multi-dimensional contextual layer. We build URL-based graphs, enrich them by assigning weights to the edges based on signals for the URLs, and generate initial campaign components within each graph (stage 1 as shown in Fig. \ref{fig:framework}). The connectivity of the graphs is determined by the associated IP addresses shared by every pair of URLs, while the signals for the URL pairs induce the edge weights. 
The generated campaign components will be further grouped into campaigns in the next section.

\subsection{Building URL Graphs}\label{sec:building url graph}

Signals collected to enrich URLs are essential to the discovery of campaigns. Among all the signals, shared IP addresses are fundamental in building the backbone of EPhishCADE, namely the URL graphs, which are then used to build the weighted graphs using other signals. An initial step at this stage involves the creation of a URL-IP bipartite graph between the collected URLs and the signal of IP addresses for those URLs. Each URL can associate with one or more IP addresses, meaning that a URL may link to more than one IP address in the bipartite graph. With the bipartite graph, we can derive URL graphs by linking two URLs if they share common IP addresses. 

\begin{definition}[URL Graph]
A URL graph $G=(U, E)$ is an undirected graph with a node set $U$ of phishing URLs where $U\subseteq \mathcal{U}$, and an edge set $E$ where each edge $e=(u_i, u_j)\in E$ indicates that $u_i$ and $u_j$ are associated with at least one common IP address. 
\end{definition}

Given a set $\mathcal{U}$ of URLs, multiple URL graphs can be generated as some URLs may not share any IP addresses with other URLs. URLs in the same graph share stronger connections and are thus more likely to have originated from the same phishing campaign. However, URL graphs alone are not enough to find campaigns. We need to incorporate more signals into the URL graphs to provide stronger connectivity for URLs in the same graph. It is also possible that URLs from the same campaign sometimes do not share common IP addresses. We will address this challenge by clustering campaign components (subgraphs from different URL graphs) based on their semantic similarity in Sec. \ref{sec:stage2}.

\subsection{Building Weighted URL Graphs} \label{sec:weighted url graph}

In this section, we incorporate other signals to enrich the URL graphs to enhance the connectivity of URLs by assigning weights to the graph edges. A signal may increase the weight of an edge depending on the signal similarity of two URLs in the edge. For instance, given a signal $s$ and an edge $e=(u_1, u_2)$ in URL graph $G$, we increase the weight $w_e$ by 1 and say that signal $s$ \textit{contributes} to edge $e$ if $s(u_1)\doteq s(u_2)$. The $\doteq$ sign means either ``equal to'' or ``highly similar to'' depending on the signal and which similarity measure is used for that signal.

Given a signal set $S$, we assess all the edges from a URL graph and turn it into a weighted graph by increasing the edge weight by 1 for each signal if the similarity of the signal values of two URLs in an edge is within a threshold $\delta$.

\begin{definition}[Weighted URL Graph]
A weighted URL graph $WG=(U,E,W)$ is a URL graph with a weight $w\in W$ assigned to each edge $e\in E$. 
\end{definition}

A weighted URL graph provides a comprehensive view of the connectivity between URLs in that graph from multiple dimensions (each signal is a dimension). If the graph is densely connected and the weights of its edges are large enough, the URLs in the graph likely belong to the same phishing campaign due to a high degree of similarity manifested in the signals. As mentioned in Sec. \ref{sec:problem}, the signals fall into two categories -- texts and comparable values. We summarize the signals we use in Table \ref{tab:Signals and Types}.

\begin{table}[t]
\centering
\caption{Signals, Types and Measurements}
\label{tab:Signals and Types}
\resizebox{0.43\textwidth}{!}{%
\begin{tabular}{c|c|c}
  \hline
  \textbf{Signal} & \textbf{Type} & \textbf{Measurement}\\ \hline
  URL Token & Textual & Similarity of Vectors\\\hline
  HTML Text & Textual & Similarity of Vectors\\ \hline
  OCR Text & Textual & Similarity of Vectors\\
  \hline
  IP Address & Comparable Value & Count\\\hline
  DNS & Comparable Value & Value\\\hline
  Reverse DNS & Comparable Value & Value\\\hline
  Target & Comparable Value & Value\\\hline
  Submission Time & Comparable Value & Difference\\\hline
\end{tabular}
}
\end{table}

\subsection{Textual-Signal Weight Updating}

Next, we explain how to update the weights of edges using textual signals first.

\subsubsection{Tokenizing URLs}
URL tokens are extracted by splitting URLs by special characters such as ``.'', ``?'' and ``-''. We observe that URLs with similar tokens often share high similarity in other signals. In a phishing campaign, attackers may modify a small part (e.g., a token) of a URL to create multiple phishing URLs to lower the chance of being identified. For instance, $u_1$ and $u_2$ in Table \ref{tab:example_malicious_urls} have the same tokens except the numbers in the first one. Thus, attackers probably launch multiple phishing attack instances by using the same domain name \texttt{paypal-login.net} but different subdomains (\texttt{s286} and \texttt{s8790}). We use the widely-used cosine similarity metric to determine if URL tokens contribute to an edge. If the similarity is larger than the threshold $\delta$, the weight of the corresponding edge increases by 1.

\subsubsection{Extracting HTML Texts}
Attackers often reuse the phishing content so that the web pages linked to different URLs may look similar. Thus, we download the web pages of the phishing URLs and use the extracted HTML texts as another signal to update edge weights. We extract web page texts without HTML encodings, convert the texts into TF-IDF vectors, and compute the cosine similarity of the vectors to find if the HTML texts of two URLs are similar. However, due to high dynamics of phishing attacks, it is not uncommon that some URLs are short-lived and the HTML texts sometimes may not show the original phishing content but error messages. To reduce noise, we develop a dictionary of commonly observed words in error pages to filter them out.  

\subsubsection{Extracting OCR Texts and Handling Information Loss}
The existence of error pages results in information loss for some URLs. To alleviate this problem, we leverage web page screenshots and extract OCR texts from screenshots as another signal to supplement HTML texts. As there are also error pages in PhishTank's screenshots, we take screenshots of web pages by ourselves. Given a URL $u$ and its two OCR texts $OCR_{our}$ and $OCR_{pt}$ extracted from screenshots of our own and PhishTank respectively, we employ the following heuristics to merge the two versions to avoid duplicates and alleviate information loss caused by missing web pages: 1) pick up the first one if the two OCR texts are the same; 2) keep the longest one if they are similar; 3) concatenate the texts if dissimilar.
Once we have the merged OCR text for each URL, we can measure the textual similarity for any pair of URLs.

\subsection{Comparable-Value Weight Updating}

As seen above, updating edge weights for URL graphs using textual signals often takes extra effort for comparison. Next, we work on signals whose values can be compared directly or through simple difference operations.

\subsubsection{Common IP Addresses Shared by URLs} Recall in Sec. \ref{sec:building url graph}, we use common IP addresses to build URL graphs. Meanwhile, we can also utilize the number of common IP addresses as another signal for edge weight update. Given two URLs $u_1$ and $u_2$ in an edge $e$, suppose that the number of common IP addresses for the two URLs is $n$ ($n\geq 1$). As $n$ could be any natural number, we increase the weight of $e$ by 1 if $n\leq \delta_{IP}$, where $\delta_{IP}$ is a threshold for the number of common IP addresses, and increase the weight by 2 if $n>\delta_{IP}$. By doing so, we can capture different magnitudes of the commonality in shared IP addresses but also avoid the dominance of the edge weights by this single signal.

\subsubsection{DNS and Reverse DNS}
The use of these two signals is straightforward. If the DNS or Reverse DNS is the same for URLs $u_1$ and $u_2$ in edge $e$, the edge weight increases by 1. Otherwise, the weight remains unchanged.

\subsubsection{Target}

Target means the brand is under attack. These values can be compared directly for each signal.

\subsubsection{Temporal Signal}
Submission time is when a phishing URL is first lodged in a threat intelligence platform. We use the submission time to approximate the actual start time of a phishing URL instance which is usually hard to find out. It is common for similar phishing attacks in a campaign to be launched close in time and last for a certain period. This enables us to measure the relevance of two attacks by the submission time of their URLs. Suppose the submission times for URLs $u_1$ and $u_2$ are $t_i$ and $t_2$ respectively. If the time difference $|t_1-t_2|$ is less than a threshold $\delta_{time}$, we consider $u_1$ and $u_2$ are similar in terms of submission time and their corresponding edge weight increases by 1. The value of $\delta_{time}$ depends on the nature of attacks in a campaign, such as the frequency of similar attacks and the duration of them. 

\subsection{Campaign Components Generation}\label{sec:generate components thru community detection}

After measuring the similarity of URL pairs in URL graphs against all signals, we turn the unweighted graphs into weighted URL graphs. These graphs capture the connectivity of phishing URLs in multiple dimensions where each signal represents a dimension. In particular, the edges indicate the connections between the URLs, while the weights reflect the ``strength'' of the connections. The higher the weights of the edges, the more likely that connected URLs are from the same campaign. Meanwhile, the edges with low weights indicate low connections of the URLs and thus a low chance of belonging to the same campaign. In light of this, we apply community detection over the weighted URL graphs to split each of them into multiple \textit{campaign components} (i.e., communities). The URLs in each component are connected with higher weighted edges and are thus more likely to come from the same campaign. Each weighted URL graph can generate at least one campaign component, i.e., the graph itself if it is indivisible. A campaign component is a subgraph of the original weighted URL graph. By assigning each campaign component an id, a weighted URL graph turns into a labeled weighted URL (LaWU) graph, with the label of a URL being the campaign component id. The campaign components of a LaWU graph $LG_i$ can be presented as $comp(LG_i)=\{cmp_i^1,cmp_i^2,...\}$, where $cmp_i^j$ is the $j$th campaign component from $LG_i$ ($j$ is also the label of the URLs in the campaign component). As an example, after community detection the weighted URL graph $WG_i$ at stage 2 in Fig. \ref{fig:framework} turns into a LaWU graph $LG_i$ having four campaign components $cmp_i^1$, $cmp_i^2$, $cmp_i^3$ and $cmp_i^4$.

\section{Campaign Generation with Components}\label{sec:stage2}

URLs in the same campaign component are connected through shared IP addresses. However, it is possible that phishing attacks in the same campaign leverage URLs that do not share common IP addresses so that their correlations will not be detected easily. 
To find out if URLs in different components belong to the same campaign, we will apply hierarchical clustering to merge similar components from across different LaWU graphs. We utilize the HTML texts with the supplement of OCR texts to measure the similarity of components. We develop two algorithms to achieve the following two goals, respectively:
\begin{itemize}
    \item{Algo 1:} Extract long docs and globally index components.
    \item{Algo 2:} Cluster components and generate campaigns.
\end{itemize}

\subsection{Extracting LongDocs and Indexing Components Globally}

Before components' clustering, we need to group the texts in each component and prepare the data. Specifically, we need to:
\begin{enumerate}
    \item concatenate the HTML texts (or OCR texts if HTML texts are unavailable) for URLs in the same component to form a single long text (LongDoc for short) for that component.
    
    \item build data structures so that searching the components and their LongDocs can be efficient. 
\end{enumerate}
Algorithm \ref{algo:community-indexing} is designed for this purpose. The algorithm takes in the LaWU graphs and produces three outputs: $long\_docs$, $gid\_for\_communities$ as well as $gid\_to\_component$. Each value of $long\_docs$ is a concatenated text (or LongDoc) using the HTML or OCR texts of URLs in a campaign component. Each LongDoc is the text content of a component and can be used to compute similarity of two components. Function \texttt{LongDoc()} in line 9 is used to generate the LongDoc given a campaign component. By default, it concatenates the HTML texts of URLs in that component. However, if it detects that an HTML text actually represent an error message, the function will use the OCR text instead. LongDocs in $long\_docs$ are indexed by a global index $gid$. The parameter $gid\_to\_component$ is a mapping from each $gid$ to a specific component in a graph. That is, given a $gid$, we can find its corresponding graph and component immediately. $gid\_for\_communities$, on the contrary, is used to find the $gid$ for a component given the index of a graph and the label of a component. These variables will be very handy in the subsequent sections when generating and accessing campaigns via hierarchical clustering.

\subsection{Clustering Components and Generating Campaigns}\label{sec:clustering components}

Given $long\_docs$, each of which is the HTML text for a campaign component, we transform them into TF-IDF vectors, which are then fit into a hierarchical clustering model to generate clusters of components. Each cluster represents a phishing campaign consisting of a number of semantically similar components. Based on the ordering in $long\_docs$, the clustering model also generates a list of labels $comp\_lbls$, with each label indicating which cluster the corresponding component belongs to. We also call the labels the campaign labels or ids. Based on the labels and their index (the global id) in $comp\_lbls$, we can use $gid\_to\_comp$ to find which components belong to a cluster with a certain label, i.e., a specific campaign. Algorithm \ref{algo:algorithm2} describes this procedure. In line 5, we obtain $i$ and $j$, which tells us that the $j$th component in the $i$th graph is labelled $lbl$. Thus, we add the corresponding campaign component $cmp_i^j$ to the list $comps\_in\_campaign$ of components for campaign $lbl$. 

\begin{algorithm}[t]
\caption{Prepare LongDocs and Reindex Campaign Components}
\begin{algorithmic}[1]
\Procedure{LongDocs}{}
\State \hspace{-5mm}\textbf{input}: A set of labeled weighted URL (LaWU) graphs $\{LG_i\}$ ($0\leq i\leq p$)
\State \hspace{-5mm}\textbf{output}: 
\State $long\_docs$: LongDocs for campaign components indexed by a global index
\State $gid\_for\_communities$: global indexes for a certain component in a graph
\State $gid\_to\_component$: a mapping from global indexes to a component in a graph
\State \hspace{-5mm}\textbf{begin}
\State \hspace{-5mm}\quad initialise $gid=0$ to be a global label index
\State \hspace{-5mm}\quad initialise $long\_docs$ to be an empty set
\State \hspace{-5mm}\quad initialise $gid\_for\_communities$ to be an empty list
\State \hspace{-5mm}\quad initialise $gid\_to\_component$ to be a dictionary
\State \hspace{-5mm}\quad \textbf{for} each graph $LG_i$ \textbf{do}
\State \hspace{-5mm}\quad\quad let $lbl\_to\_gid$ be a dictionary mapping a component label to a global label index
\State \hspace{-5mm}\quad\quad \textbf{for} each campaign component $cmp_i^j$ in $LG_i$ \textbf{do}
\State \hspace{-5mm}\quad\quad\quad  let $long\_docs[gid]$ = getLongDoc($cmp_i^j$)
\State \hspace{-5mm}\quad\quad\quad  let $lbl\_to\_gid[j]=gid$
\State \hspace{-5mm}\quad\quad\quad  let $gid\_to\_component[gid]=(i,j)$
\State \hspace{-5mm}\quad\quad let $gid\_for\_communities[i]=lbl\_to\_gid$
\State \hspace{-5mm}\quad \textbf{return} $long\_docs$,
\State \quad\quad\quad $gid\_for\_communities$,
\State \quad\quad\quad $gid\_to\_component$
\EndProcedure
\end{algorithmic}
\label{algo:community-indexing}
\end{algorithm}

\begin{algorithm}[t]
\caption{Find components for a campaign (cluster)}
\begin{algorithmic}[1]
\Procedure{Cluster}{}
\State \hspace{-5mm}\textbf{input}:  A mapping $gid\_to\_component$ from global indexes to components in LaWU graphs obtained in Algorithm 1;
        a list $comp\_lbls$ of labels for all campaign components;
        a campaign label $lbl$
\State \hspace{-5mm}\textbf{output}: A list $comps\_in\_campaign$ of campaign components belonging to a campaign labelled by $lbl$
\State \hspace{-5mm}\textbf{begin}
\State \hspace{-5mm}\quad initialise $comps\_in\_campaign$ to be an empty list
\State \hspace{-5mm}\quad let $indexes\_lbl$ be a list of indexes of all occurrences of label $lbl$ in $comp\_lbls$
\State \hspace{-5mm}\quad \textbf{for} each $gid$ in $indexes\_lbl$ \textbf{do}
\State \hspace{-5mm}\quad\quad $i,j$ = $gid\_to\_component[gid]$
\State \hspace{-5mm}\quad\quad add $cmp_i^j$ to $comps\_in\_campaign$
\State \hspace{-5mm}\quad \textbf{return} $comps\_in\_campaign$
\EndProcedure
\end{algorithmic}
\label{algo:algorithm2}
\end{algorithm}

%% file: experiment1.tex
\begin{table*}[th]
\centering
\caption{Signal Strengths for Campaigns}
\label{tab: Signal Strengths}
\begin{tabular}{c|c|l}
  \hline
  Campaign & Avg SigS & Signals Sorted based on Individual SigS \\ 
  \hline \hline
  1 & 0.4912 & geoip:1.0, reversedns:1.0, target:0.89, dns: 0.78, countrycode:0.73, submission\_time:0.4708, ocr:0.2789\\ 
  \hline
  2 & 0.4151 & geoip:1.0, countrycode:1.0, reversedns:1.0, submission\_time:0.58, url:0.26, dns:0.25, ocr:0.12\\ 
  \hline
  3 & 0.4098 & geoip:1.0, countrycode:1.0, reversedns:1.0, dns:0.61, submission\_time:0.43, ocr:0.04, target:0.04\\ 
  \hline
  4 & 0.3991 & geoip:1.0, countrycode:1.0, reversedns:1.0, dns:0.55, submission\_time:0.37, url:0.04, ocr:0.02\\ 
  \hline
  5 & 0.3776 & geoip:1.0, reversedns:1.0, submission\_time:0.86, dns:0.73, ocr:0.03, url:0.01,  html\_tags:0.006\\ 
  \hline
\end{tabular}
\end{table*}

\section{Experimental Study}\label{sec:experiment}

In this section, we conducted an experimental study to assess the performance of our model. We first evaluate the efficacy of our proposed contextual layer using two metrics. Next, we holistically benchmarked our framework against the single-dimension structure-based study \cite{DOMCluster} (WWW \textquotesingle 17), which serves as a baseline for grouping phishing attacks based on similar DOM structures. We implemented this structure-based approach and incorporated it as the structural layer in our framework.

\subsection{Contextual Layer}

\subsubsection{Evaluation Metrics}

To evaluate the quality of the discovered phishing campaigns, we propose two metrics for this purpose -- signal strength (SigS) and coherence map (CohM). Next, we will introduce these metrics and demonstrate how one can use them to assess phishing campaigns.

\begin{figure*}[!h]
\centering 
\begin{subfigure}[t]{0.41\textwidth}
\includegraphics[width=1\linewidth]{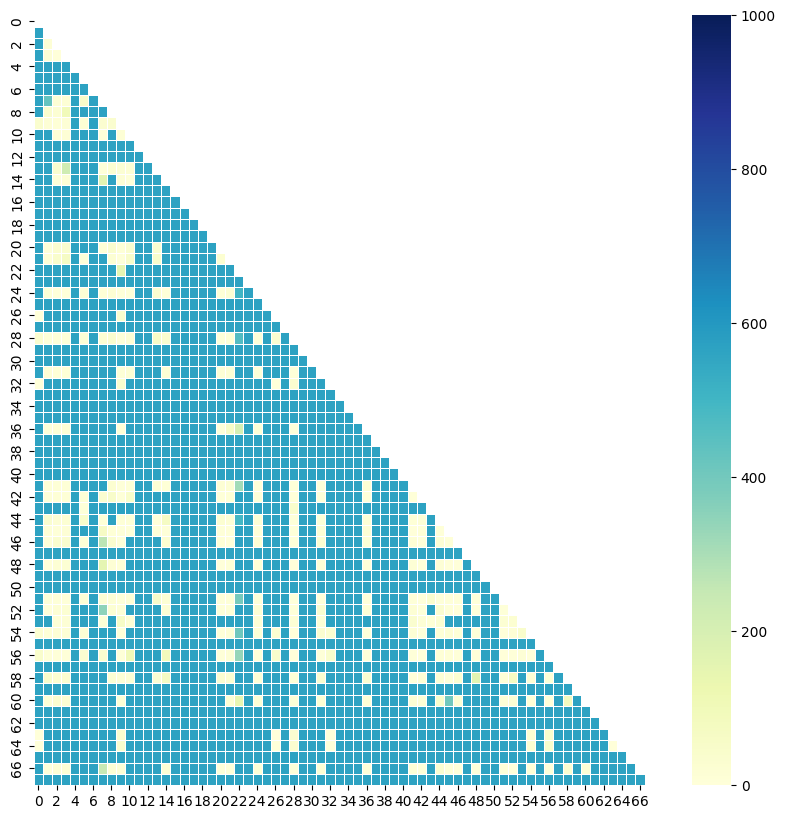}
\caption{Dataset 1 with 68 Campaigns.}
\label{fig:coh map 5k-1000}
\end{subfigure} \quad \quad \quad \quad 
\begin{subfigure}[t]{0.4\textwidth}
\includegraphics[width=1\linewidth]{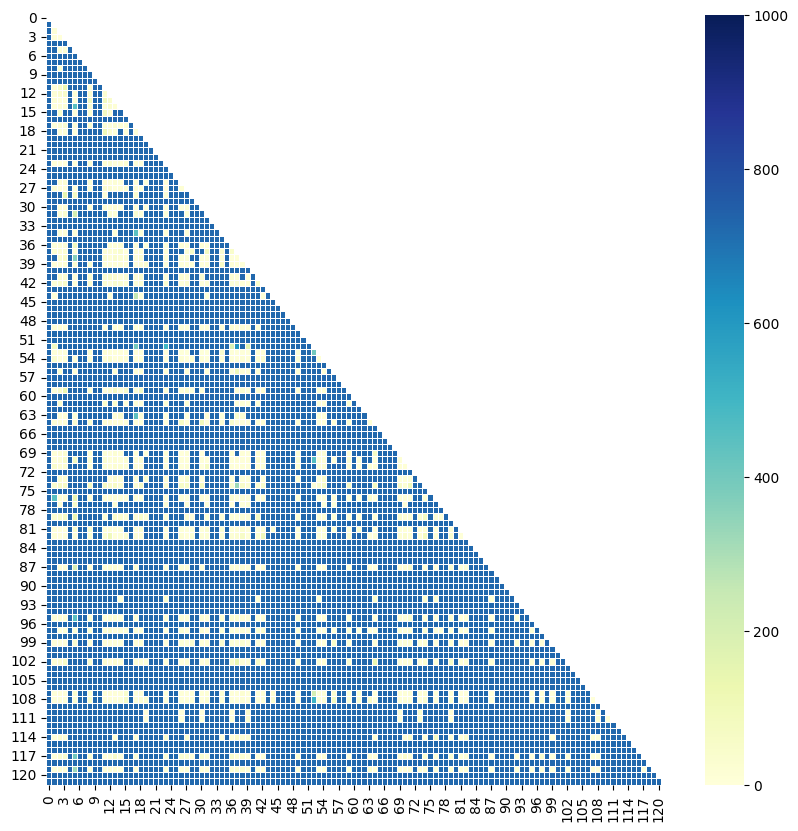}
\caption{Dataset 2 with 122 Campaigns.}
\label{fig:coh map 10k-1000}
\end{subfigure} 
\caption{Coherence Maps for Campaigns in Datasets 1 \& 2.}
\label{fig:coherence maps}
\end{figure*}

\smallskip
\noindent
{\bf Signal Strength.} Recall that we collect various signals from third parties to enrich the URLs in the data enrichment step and then generate campaigns through URLs graphs and hierarchical clustering based on the signals. However, it is important to note that not all signals contribute to the discovery of campaigns equally. That is, some signals tend to be more important than others during the entire process, and thus need to be regarded as  indispensable when generating campaigns. We estimate their importance by calculating the \textit{signal strength} (SigS), which provide a means to rank different signals and pick up the most important ones for campaign discovery.

Given a weighted graph $G_i$ ($i\in I$) and a signal $s_j$ ($j\in J$), we use $c_{inc}^{i,j}$ and $c_{all}^{i,j}$ to denote the number of edges that signal $s_j$ contributes to (i.e., the number of edges whose weights have been increased by signal $s_j$) and the number of total edges in $G_i$. Thus the signal strength of signal $s_j$ in $G_i$ can be computed as $\theta^{i,j}={cnt_{inc}^{i,j} / cnt_{all}^{i,j}}$. For a campaign $c_k$ ($k\in K$) with $M$ components $\{cmp_{(k)(1)},cmp_{(k)(2)},...,cmp_{(k)(M)}\}$, the average signal strength of $j$th signal for campaign $c_k$ is computed as 

\begin{equation}
SigS(k,j)={1 \over M}  \sum_{i\in I} (\sum_{m\in M,cmp_{(k)(m)}\subseteq G_i}{\theta_{i,j}}) 
\end{equation}
where $cmp_{(k)(m)}$ is the $m$th component in campaign $c_k$ and $cmp_{(k)(m)}\subseteq G_i$ means component $cmp_{(k)(m)}$ comes from Graph $G_i$ through Section \ref{sec:generate components thru community detection}. Note not to confuse $cmp_{(k)(m)}$ (a component at the campaign level) with $cmp_i^j$ (a component at the graph level) mentioned in Section \ref{sec:clustering components}.

\begin{table*}[!h]
\caption{A detailed numeric comparison of the structured-based solution and our work.}
\resizebox{0.98\textwidth}{!}{%
\begin{tabular}{|c|c|c|cccccc|}
\hline
\textbf{Type}            & \textbf{No. of Campaigns} & \textbf{\begin{tabular}[c]{@{}c@{}}No. of Single-URL Campaign\\ (Individual Attack)\end{tabular}} & \multicolumn{6}{c|}{\textbf{No. of Multi-URL Campaigns}}                                                                                                                          \\ \hline
\multirow{2}{*}{-}       & \multirow{2}{*}{-}        & \multirow{2}{*}{-}                                                                                & \multicolumn{4}{c|}{$\leq 5$ URLs}                                                                                        & \multicolumn{1}{c|}{$> 5$ URLs} & Total     \\ \cline{4-9} 
                         &                           &                                                                                                   & \multicolumn{1}{c|}{2 URLs}    & \multicolumn{1}{c|}{3 URLs}    & \multicolumn{1}{c|}{4 URLs}   & \multicolumn{1}{c|}{5 URLs}   & \multicolumn{1}{c|}{-}                   & -         \\ \hline
\textbf{Structure-Based} & 1050                      & 615                                                                                               & \multicolumn{1}{c|}{192}       & \multicolumn{1}{c|}{66}        & \multicolumn{1}{c|}{37}       & \multicolumn{1}{c|}{28}       & \multicolumn{1}{c|}{112}                 & 435       \\ \hline
\textbf{Our Work}        & 891                       & 552                                                                                               & \multicolumn{1}{c|}{169}       & \multicolumn{1}{c|}{52}        & \multicolumn{1}{c|}{30}       & \multicolumn{1}{c|}{20}       & \multicolumn{1}{c|}{68}                  & 339       \\ \hline
\textbf{Reduction}       & 159 (15\%)                & 63 (10\%)                                                                                         & \multicolumn{1}{c|}{23 (12\%)} & \multicolumn{1}{c|}{14 (21\%)} & \multicolumn{1}{c|}{7 (19\%)} & \multicolumn{1}{c|}{8 (29\%)} & \multicolumn{1}{c|}{44 (39\%)}           & 96 (22\%) \\ \hline
\end{tabular}
}
\label{tab:comparison}
\end{table*}

Below, we use around 20.1k URLs collected between \texttt{2023-08-15} and \texttt{2023-09-14} and the enrichment data to generate 114 campaigns with signal strengths. In Table \ref{tab: Signal Strengths}, we only display 5 campaigns with the highest average signal strength. The third column is a sorted list of (top 7) signals for each campaign based on the individual signal strength. It is worth noting that the average signal strength varies across campaigns and that the signal strength of each individual signal may also vary in different campaigns. This tells us that the importance of signals can be different in different campaigns. For instance, while \texttt{countrycode} is one of the most important signals for campaigns 2, 3 and 4 as the signal strength is 1.0, it does not even appear among the top 7 signals for campaign 5. Another observation is that if a signal appears as the most important one across all the campaigns (such as \texttt{geoip}), it may however lose the capability of distinguishing different campaigns during the discovery process. Therefore, this type of signal can be removed without affecting the discovery outcome.

Signal strength can be used to select important signals for the entire campaign discovery process and for finding out which signals play an important role in each individual campaign.

\smallskip
\noindent
{\bf Coherence Map.} We also develop another metric named \textit{coherence map} (CohMap) to measure the distance of any pair of discovered campaigns. Ideally, the URLs and the corresponding signals in the same campaign need to be as similar as possible while those in different campaigns need to be as dissimilar as possible.

Based on this idea, we calculate a \textit{coherence score} $Coh(c_i, c_j)$ for two campaigns $c_i$ and $c_j$ by estimating an intra-campaign similarity and an inter-campaign similarity and combining them together as follows

\begin{equation}
\label{eq:coherence score}
Coh(c_i, c_j)= {{{intra\_campgn\_sim(c_i)+intra\_campgn\_sim(c_j)} \over 2} \over {inter\_campgn\_sim(c_i,c_j)}}
\end{equation}
where $intra\_campgn\_sim(c_i)$ and $intra\_campgn\_sim(c_j)$ are the intra-campaign similarity for $c_i$ and $c_j$ respectively and $inter\_campgn\_sim(c_i,c_j)$ is the inter-campaign similarity for the pair of $c_i$ and $c_j$. The former similarity measures the similarity of components within a campaign while the latter measures the similarity between campaigns. To compute the intra-campaign similarity for campaign $c_i$, we convert all the components in the campaign into TFIDF vectors, compute the cosine similarity for every pair of vectors and take the average over the similarities. To compute the inter-campaign similarity for two campaigns $c_i$ and $c_j$, we merge all the components in each campaign to form a larger vector for the campaign, and compute the cosine similarity using the larger vectors for both campaigns.

By combining the two types of similarity metrics together as in Eq. \ref{eq:coherence score}, we compute the coherence score for two campaigns to measure the distinguishability of two campaigns. To achieve high distinguishability, the intra-campaign similarity needs to be large while the inter-campaign similarity needs to be small. As both similarity metrics range between 0 and 1, the coherence score needs to be at least larger than 1 to achieve a relatively high distinguishability. With its value less than 1, we consider the two campaigns achieve a poor distinguishability.

\begin{figure*}[!h]
\centering 
\begin{subfigure}[t]{0.48\textwidth}
\includegraphics[width=1\linewidth]{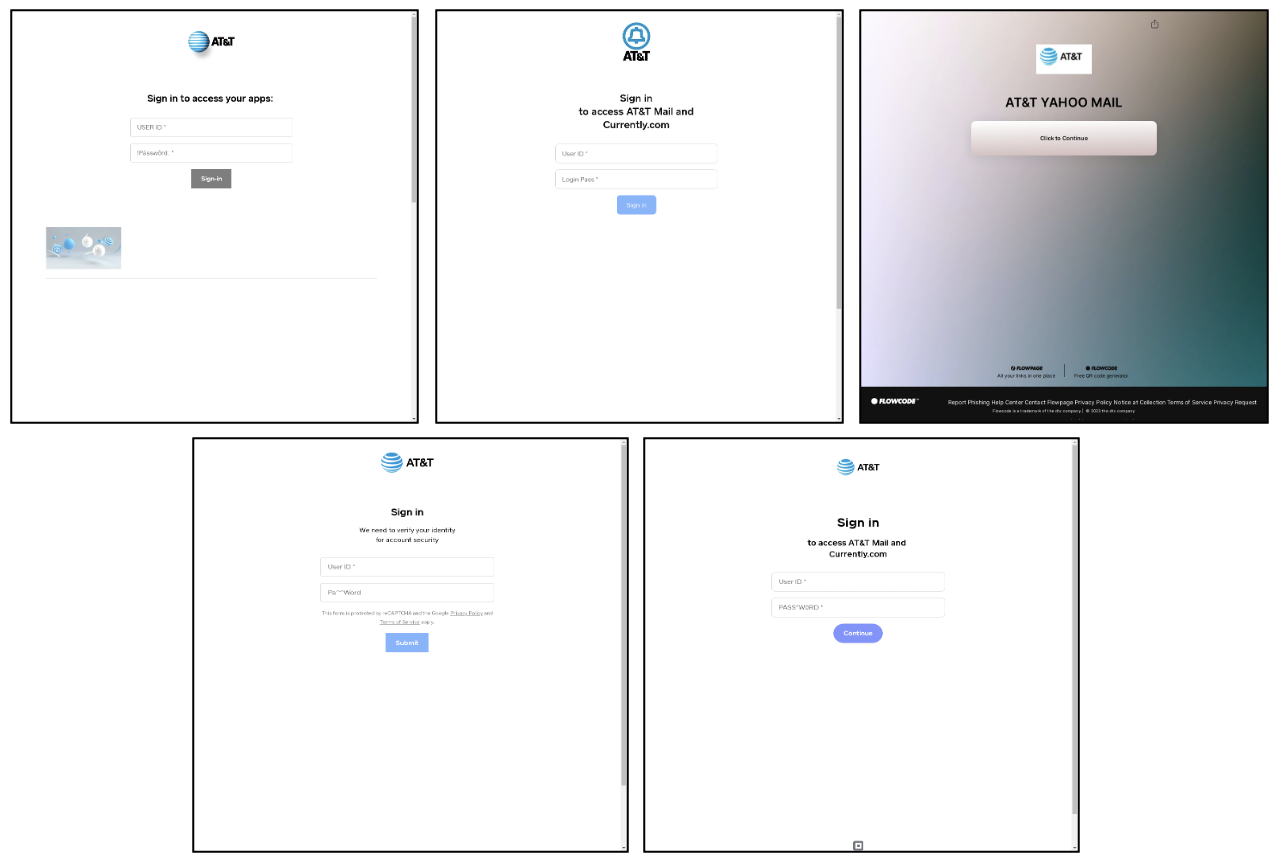}
\caption{Case I.}
\label{fig:case1}
\end{subfigure} 
\begin{subfigure}[t]{0.48\textwidth}
\includegraphics[width=1\linewidth]{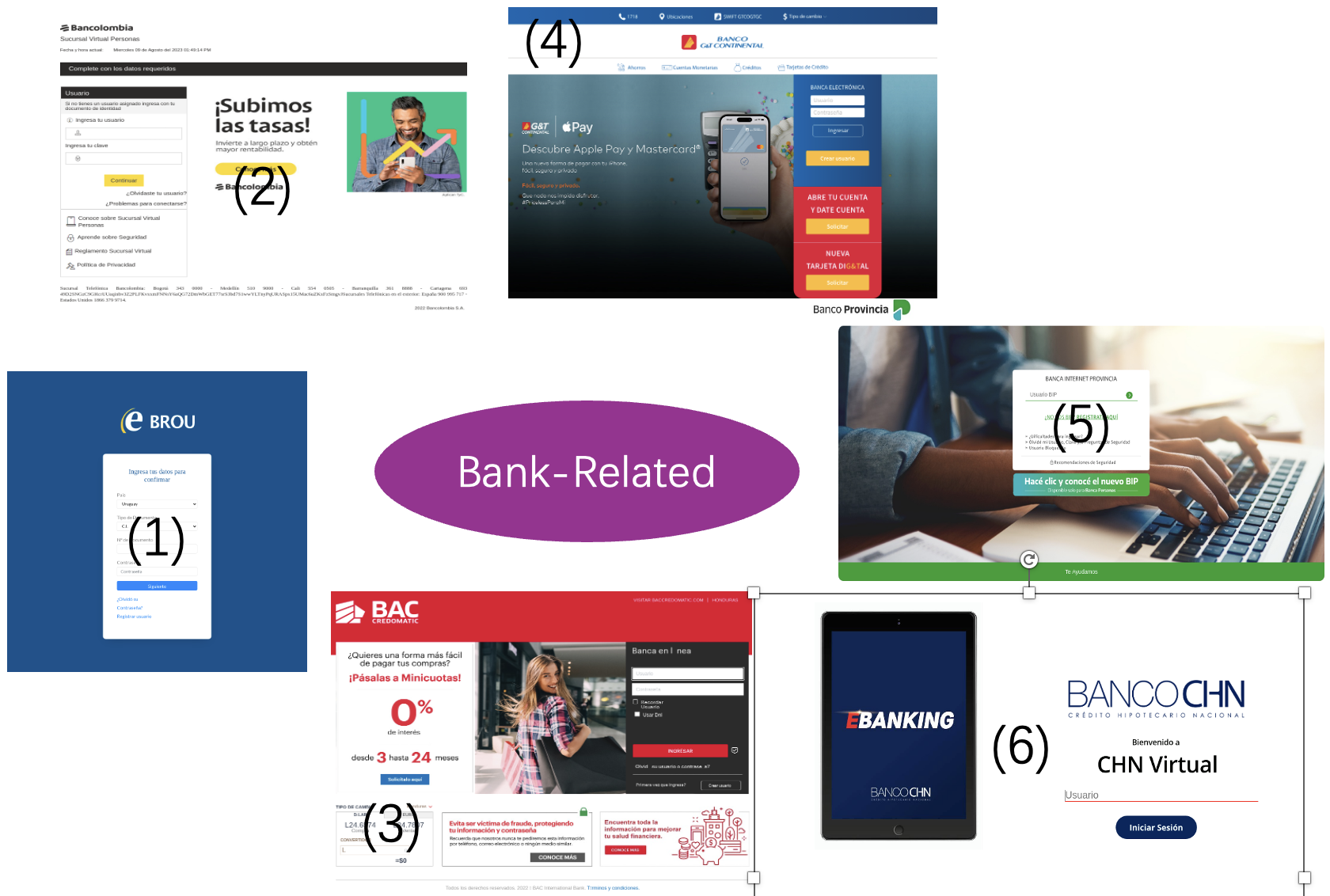}
\caption{Case II.}
\label{fig:case2}
\end{subfigure}
\caption{Contextual Layer: Two Illustrative Examples Built on Structural Layer}
\end{figure*}

\begin{figure*}[!h]
\centering
\includegraphics[width=0.65\textwidth]{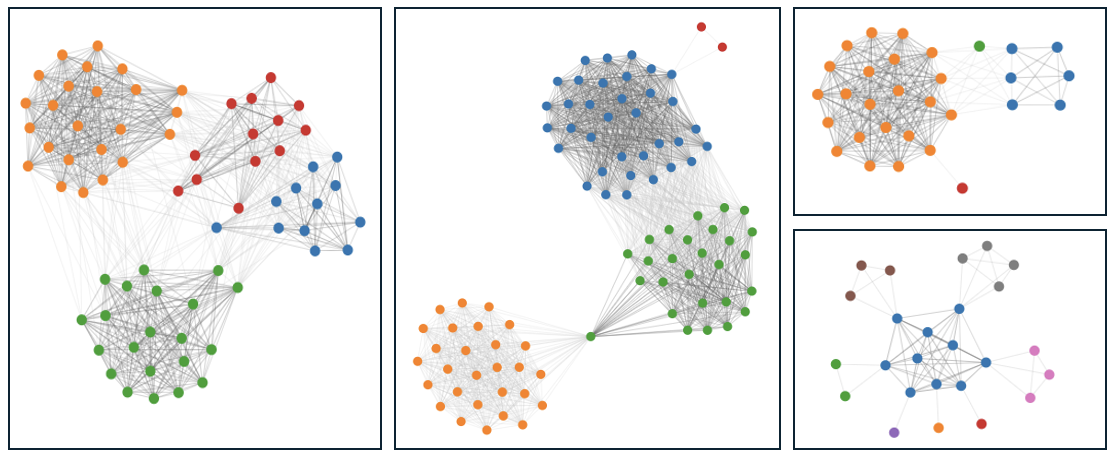}
\caption{The weighted URL graphs of campaigns.}
\label{fig:graphs}
\end{figure*}

By calculating the coherence score for every pair of campaigns, we can produce a coherence map (CohMap), which is a heatmap of coherence scores for all the campaigns. Fig. \ref{fig:coherence maps} shows two CohMaps generated for Datasets 1 and 2 respectively. Each cell in the CohMaps corresponds to a coherence score for a pair of campaigns with the color indicating the value of the score. Fig. \ref{fig:coh map 5k-1000} indicates that the scores are higher than $500$ (cells with the sky blue color) for most ($83.7\%$) coherence scores while Fig. \ref{fig:coh map 10k-1000} also show similar value distribution ($83.7\%$ scores above $720$), which means that the discovered campaigns for both datasets have high distinguishability, In other words, the similarity of communities in the same campaign is often high and that of different campaigns is usually low.

%% file: experiment.tex
\subsection{Combination of Two Layers} 

In this section, we collected 4,664 verified URLs from the well-known third-party threat intelligence platform PhishTank between August 1, 2023, and August 17, 2023, and proceeded to enrich them. After filtering out those that led to error pages, we built a dataset of 3,812 phishing URLs and their corresponding enrichment data.

\subsubsection{Numeric Comparison}

Table \ref{tab:comparison} presents a detailed numeric comparison between the structure-based solution and our proposed framework. As a key objective, our framework seeks to alleviate security analysts' workload and enhance processing efficiency. The results demonstrate that our framework reduces the total number of campaigns identified by solely structured-based approach from 1050 to 891, achieving a notable 15\% reduction. Particularly, within the 1050 campaigns, approximately 60\% consist of only a single URL, indicating isolated attacks. Our framework achieves a 10\% reduction in this proportion, demonstrating its ability to uncover hidden connections among seemingly independent phishing attacks. Additionally, for campaigns involving multiple ($>1$) URLs, its performance improves significantly, with a 22\% reduction. Notably, for campaigns with $\leq 5$ URLs and $> 5$ URLs, our framework achieves satisfactory reductions, respectively. This demonstrates its capacity to consolidate related phishing instances, allowing analysts to focus on meaningful patterns and reducing redundant efforts, ultimately leading to more efficient and comprehensive phishing detection and response.

\subsubsection{Case Studies}

We provide two representative case studies illustrated in Figure \ref{fig:case1} and \ref{fig:case2}, respectively.

\smallskip
\noindent
{\bf Case I.} The structural layer successfully identified five distinct groups of URLs targeting the well-known AT\&T brand. However, these URLs exhibited variations in their DOM structures, resulting in difficulties for structure-based solutions to detect their semantic connections. Our contextual layer was able to further analyze additional features such as content, behavior, and metadata. This allowed the framework to effectively group these seemingly different phishing URLs into a unified cluster, demonstrating its ability to detect and correlate phishing attacks that employ subtle structural differences to evade detection. The combination of the structural and contextual layers thus enhances the framework's ability to capture sophisticated phishing campaigns targeting the same entity.

\smallskip
\noindent
{\bf Case II.} The structural layer identified six distinct groups of URLs, each with completely different DOM structures. However, our contextual layer was able to uncover latent relationships among these groups that were not immediately apparent from their structural analysis. All six groups were associated with banking activities. Specifically, the group (1) was also linked to the three groups (2), (3), (4) through a shared IP address, suggesting the possibility of a common attacker orchestrating multiple phishing campaigns. This deeper contextual insight enables security analysts to conduct more thorough investigations, connecting seemingly unrelated attacks and flagging potential coordinated threats. As a result, the system can generate timely alerts for further scrutiny and defense actions, helping to mitigate widespread phishing activities.

\subsubsection{Weighted URL Graphs}

We developed a web-based campaign monitoring tool that visually presents the weighted URL graphs of identified campaigns. Figure \ref{fig:graphs} displays the weighted URL graphs of four distinct campaigns, where each node represents a URL and the edges denote relationships between URLs. The colors signify coarse groups identified by the structural layer. It can be seen that our contextual layer further consolidates the groups formed by the structural layer.

%% file: conclusion.tex
\section{Conclusion}

In this paper, we introduced EPhishCADE, a novel privacy-aware, multi-dimensional framework designed to automatically identify phishing email campaigns by clustering seemingly unrelated attacks. EPhishCADE employs an hierarchical architecture that combines both structural and contextual layers, enabling a comprehensive analysis of phishing activity by thoroughly examining these elements. Specifically, the framework’s graph-based contextual layer reveals hidden correlations across multiple dimensions between attacks that may initially appear disconnected. Importantly, our framework relies solely on URL-related data, avoiding the use of sensitive email content to maintain privacy. EPhishCADE improves users' ability to detect phishing campaigns, alleviating alert fatigue for security analysts and organizations, while helping them prioritize resources for the most critical threats. 